%% file: 0_PP_keystroke.tex
\documentclass{article}

\usepackage{PRIMEarxiv}
\usepackage{epsfig}
\usepackage[utf8]{inputenc} 
\usepackage[T1]{fontenc}    
\usepackage{hyperref}       
\usepackage{url} 
\usepackage{comment}
\usepackage{booktabs}       
\usepackage{amsfonts}       
\usepackage{nicefrac}       
\usepackage{microtype}      
\usepackage{lipsum}
\usepackage{fancyhdr}       
\usepackage{graphicx}       
\graphicspath{{media/}}     
\usepackage{algorithm}
\usepackage{algorithmic}
\usepackage[square,sort,comma,numbers]{natbib}
\usepackage{lipsum}	
\usepackage{multirow}
\usepackage{url}            
\usepackage{booktabs}
\usepackage{epsfig}
\usepackage{subcaption}
\usepackage{calc}
\usepackage{amssymb}
\usepackage{amstext}
\usepackage{amsmath}
\usepackage{amsthm}
\usepackage{tikz}
\usepackage{caption}
\usepackage{natbib}
\title{A Generic Privacy-Preserving Protocol For Keystroke Dynamics-Based Continuous Authentication
\thanks{This manuscript is an updated version of the following paper: \\
\textit{\underline{Citation}}: 
\textbf{Baig, A. and Eskeland, S.
A Generic Privacy-preserving Protocol for Keystroke Dynamics-based Continuous Authentication.
DOI: 10.5220/0011141400003283
In Proceedings of the 19th International Conference on Security and Cryptography (SECRYPT 2022), pages 491-498
ISBN: 978-989-758-590-6; ISSN: 2184-7711 (\cite{secrypt22})
}} 
}

\author{
   Ahmed Fraz Baig and Sigurd Eskeland\\
 Norwegian Computing Center \\ Postboks 114 Blindern \\ 0314 Oslo, Norway
  \texttt{\{baig,sigurd\}@nr.no} 
}

\begin{document}


\maketitle

\begin{abstract}
Continuous authentication utilizes automatic recognition of certain user features for seamless and passive authentication without requiring user attention.
Such features can be divided into categories of physiological biometrics and behavioral biometrics.
Keystroke dynamics is proposed for behavioral biometrics-oriented authentication by recognizing users by means of their typing patterns.
However, it has been pointed out that continuous authentication using physiological biometrics and behavior biometrics incur privacy risks, revealing personal characteristics and activities.
In this paper, we consider a previously proposed keystroke dynamics-based authentication scheme that has no privacy-preserving properties.
In this regard, we propose a generic privacy-preserving version of this authentication scheme in which all user features are encrypted --- preventing disclosure of those to the authentication server.
Our scheme is generic in the sense that it assumes homomorphic cryptographic primitives.
Authentication is conducted on the basis of encrypted data due to the homomorphic cryptographic properties of our protocol.
\end{abstract}
\keywords{Continuous Authentication, Homomorphic Encryption, Keystroke dynamics, Behavioral biometrics.}

\input{1_intro}

\input{2_related}

\input{4_preliminaries}

\input{5_HE-contiouos}
\input{7_analysis}
\input{8_conclusion}


\section*{\uppercase{Acknowledgement}}
{This work is part of the Privacy Matters (PriMa) project. The PriMa project has received funding from European Union’s Horizon 2020 research and innovation programme under the Marie Skłodowska-Curie grant agreement No.~860315.
The authors would like to thank Dr. Wolfgang Leister for valuable comments.}

\bibliographystyle{apalike}
\bibliography{report}  

\end{document}

%% file: 1_intro.tex
\section{\uppercase{introduction}} \label{sec:intro}
User authentication is a process that confirms whether a user is the one who he claims to be.
The most common form of authentication is session-oriented authentication, where a session has a certain duration, and the user authenticates himself once at the start of a session.
Such authentication mechanisms are mainly classified into the following categories:
 Knowledge-based authentication (what you know, like passwords and PIN codes),
   possession-based authentication (what you have, such as smartcards or dongles)
   physiological biometrics (face recognition, iris recognition, fingerprint recognition, etc.).
Session-orientation implies that the user is required to do some active or explicit action up front, like typing a password, inserting a smartcard, or scanning his fingerprint.
Session-oriented authentication approaches authenticate users at the beginning of the session.
If the user leaves the device for some time, the device will remain unlocked for a time, which could allow a malicious user to use the device in the meantime.

For computer devices that are carried by humans, such as smartphones, continuous authentication has been proposed to strengthen the mentioned authentication methods.
The supposed advantage is \emph{passive and seamless} authentication mechanisms that do not require user attention.
The idea of continuous authentication is that there is some uniqueness to some user biometry or user context.
The authentication process is automatically conducted by events of relevant user activity.
The time window of access is much smaller than for session-oriented approaches, and the system automatically locks in case the user is inactive or when it observes anomalous behaviors.

Continuous authentication can be achieved by following categories of modes: Behavioral biometrics, physiological biometrics, and context-aware authentication.
The overall premise for behavioral biometrics is that every person has a uniqueness in walking style, typing style, movement, and so on.
By recognizing such movement patterns, a person can be uniquely identified.
Behavioral biometric modes include touch screen dynamics, keystroke dynamics, stylometry, gait and walking style, etc.

Physiological biometrics, including face and iris recognition, are often considered for continuous authentication, although such modalities normally require explicit user actions and fails as such being passive and seamless.

Context-aware authentication modes include IP-addresses, operating systems, and other profiling parameters such as GPS position, battery usage, network usage, web browsing histories, and other behavioral activities.
Context-aware modes rely on contextual parameters of the device, such as IP addresses, location data, etc. The potential problem with context-aware modes is that they only re-authenticate users when there is a change in contextual information. But they cannot differentiate between a legitimate user or an imposter, if there is no contextual change. Such problems may occur when users leave their devices open, and someone else uses their devices in their absence.

The definition of continuous authentication demands that the selected mode needs to be passive and continuous simultaneously. Therefore, the physiological biometrics and context-aware modes cannot be solely considered for continuous authentication. Only behavioral biometrics fulfill the requirements of continuous authentication, due to their passive and continuous nature \citep{baig2021security}.

Keystroke dynamics are categorized as behavioral biometrics that authenticates users by analyzing and recognizing user typing behaviors and typing patterns.
The keystroke dynamics authentication mechanism can be implemented either in continuous way, where the user is identified on each input \citep{bours2012continuous} or in periodic way, where user validity is confirmed over a collected block of actions; the decision is based on the analysis of that block of data \citep{dhakal2018observations,xiaofeng2019continuous}.
A minor disadvantage of periodic authentication is the delay for the authentication decision to take place, while for continuous authentication this is conducted immediately at every user event \citep{bours2012continuous}.

The problem about continuous authentication methods including behavioral modalities is that there is no privacy protection.
The behavioral features of keystroke dynamics are privacy sensitive, and may disclose sensitive user information related to gender, age, left-or right-handedness, and even emotional states during typing \citep{brizan2015utilizing}.
Behavioral biometrics data are categorized as sensitive data in GDPR, Article~4.

In this paper, we propose a privacy-preserving protocol that is based on the \citet{bours2012continuous} continuous authentication scheme.
To mitigate privacy issues, our protocol uses generic homomorphic cryptographic methods; this enables the authentication operations to be conducted in the encrypted domain.

%% file: 2_related.tex
\section{\uppercase{related work}} \label{sec:related}

%
%
\citet{govindarajan:2013} proposed a \emph{periodic} privacy-preserving protocol for touch dynamics-based authentication.
Their scheme utilizes private comparison protocol proposed by \citet{erkin:2009} and the homomorphic DGK encryption algorithm proposed by \citet{damgaard:2008}.
Note that the \citet{erkin:2009} comparison protocol is based on the private comparison protocol proposed by \citet{damgaard:2007,damgaard:2009}.
The scheme of Govindarajan et al. does not reveal anything, because it makes comparisons in the encrypted domain.
However, it is not efficient for continuous authentication, mainly because of the inefficiency of the Erkin et al. subprotocol, which requires that each bit of the inputs are encrypted.
In the protocol, each of these ciphertexts are then sent to the other party.

\citet{balagani:2018} proposed a keystroke dynamics-based privacy-preserving authentication scheme.
They extended the idea of \citeauthor{govindarajan:2013}\space protocol, but is also based on the private comparison protocol proposed by \citet{erkin:2009} and the homomorphic DGK encryption algorithm proposed by \citet{damgaard:2008}.
This scheme has the same efficiency problems as the scheme by \citeauthor{govindarajan:2013}\space

\citet{wei:2020} proposed a privacy-preserving authentication scheme for touch dynamics using homomorphic encryption properties.
It is based on similarity scores between input and reference features using cosine similarity.
The authentication server performs a comparison between the encrypted reference template (provided during enrollment) and encrypted input template sampled during authentication.
The authentication server decrypts the similarity scores and compares them with a predefined threshold.

\citet{safa:2014} proposed a privacy-preserving generic protocol by utilizing context-aware data features such as users GPS data, search histories (cookies), etc. Additive homomorphic encryption properties and order-preserving symmetric encryption (OPE) are utilized to achieve the privacy of users data features. Their protocol uses the Average Absolute Deviation (AAD) for the comparison between input feature and the reference features during the authentication phase.

\citet{shahandashti:2015} proposed an implicit authentication scheme by utilizing order-preserving symmetric encryption (OPSE) with additive homomorphic encryption.
The primitives are generic, but the authors suggest the OPSE scheme proposed by \citet{boldyreva:2009} and the Paillier public key scheme.
They consider different features for implicit authentication such as user location, visited websites, etc.
Further, the AAD is utilized to compute the similarity between input and reference templates.  

\citet{domingo:2015} proposed a privacy-preserving authentication scheme using generic features, such as device data, carrier data, location data, user data stored in the cloud, etc.
They utilize set intersection to determine the dissimilarity between reference data and input data.
The privacy is protected by means of the Paillier cryptosystem and a private set intersection computation protocol proposed by \citet{freedman:2004}.
However, a potential problem with these protocols~\citep{domingo:2015,shahandashti:2015} is that context-aware modes cannot tell whether the user is present or not. Thus, if the device is stolen within the specified domain, it cannot distinguish between a legitimate user and imposters \citep{baig2021security}.

%% file: 4_preliminaries.tex
\section{\uppercase{The Bours continuous authentication scheme}} \label{sec:ca}

This section revisits the keystroke dynamics-based continuous authentication scheme proposed by \citet{bours2012continuous}.
The Bours authentication scheme is shown in Algorithm~\ref{fig:trustaggregation}, and it consists of two phases: An enrollment phase and an authentication phase, that are presented next.

\vspace{2mm}

\noindent \textbf{Enrollment phase.}
Keystroke dynamics-based authentication schemes utilize time-related data from a keystroke. The timing data are extracted in the form of features when a key is pressed down $(t_i^{down})$ and when the key is lifted up $(t_i^{up})$.
The time difference $t_i= t_i^{up}-t_i^{down}$ is computed for each keystroke, where $i$ is the index $i^{th}$ key, such as  'A' is $i=0$, 'B' is $i=1$, etc.
Based on $t_i$, further statistical analysis is performed by computing the mean $\mu_i$ and the standard deviation $\sigma_i$ for each key.
Finally, a reference template is created, which contain the statistical values $(\mu_i, \sigma_i)$ for each key.
These reference templates are then stored in the database for the purpose of authentication.

\vspace{2mm}

\noindent \textbf{Authentication phase.}
In this phase, an input template is sampled for subsequent comparison with the prestored reference template.
The authentication phase continuously takes the sampled time difference $t_i$ and computes Scaled Manhattan Distance (SMD) between $t_i$ and the reference template~$(\mu_i, \sigma_i)$~according~to
\begin{equation} \label{eq:smd}
d_i =  \frac{|t_i-\mu_i|}{\sigma_i}
\end{equation}
The distance $d_i$ is compared to the predefined threshold $T_{\text{dist}}$ in order to update the aggregated distance indicator ($C$), which is increased or decreased based on distance $d_i$.
A small value of $d_i$, close to zero, indicates the similarity between the input and the reference templates, while a greater value $d_i$ indicates dissimilarity.
Initially, $C$ is assigned a maximum value \emph{max}.

When $d_i>T_{dist}$, then $C$ is decremented in the form of a penalty function $C\gets(C-d_i+T_{\text{dist}})$.
Otherwise, $C$ is incremented in form of a reward function $C \gets min(C+R, \text{max})$ by $R$ to at most \emph{max}, where $R$ is a constant reward value.
Note that $C$ cannot exceed the maximum value.
The $C$ is continuously compared to the reject threshold $T_{\text{reject}}$ on each input.
When $C$ goes below the reject threshold ($C<T_{\text{reject}})$, the authentication fails and then the user is rejected.

\begin{algorithm}[t]
\caption{Bours keystroke dynamics-based authentication scheme}
\begin{algorithmic}
\STATE \textbf{Enrollment phase }
\STATE Compute $\mu_i, \sigma_i$, $1 \leq i \leq n$
\STATE
\STATE \textbf{Authentication phase}
\STATE C = \text{max}
\WHILE{IsKeyReleased($i$)}
\STATE  $t_i= t_i^{up}-t_i^{down}$
\STATE $d_i =  |t_i-\mu_i|/\sigma_i$ 
\IF {$d_i>T_{\text{dist}}$}
\STATE $C \gets (C-d_i+T_{\text{dist}})$   // Penalty function
\ELSE
\STATE{$C \gets min(C+R, \text{max})$}   // Reward function
\ENDIF
\IF{$C<T_{\text{reject}}$}
\STATE Reject
\ENDIF
\ENDWHILE
\end{algorithmic}
\label{fig:trustaggregation}
\end{algorithm}

%
%
%
%
\subsection{Adversarial model}\label{sec:prel1}

We assume that authentication server is semi-honest adversary, that will not deviate from the defined protocol but will attempt to learn all possible information from legitimately received messages.
The privacy requirement is that the stored reference templates and input templates are protected so that the server cannot learn anything about them.
We assume that the communication between the user and the server is secure, and that external threats are mitigated by applying network security techniques.



%% file: 5_HE-contiouos.tex

\section{\uppercase{Proposed protocol}} \label{sec:ca1}

In this section, we present a new generic privacy-preserving continuous authentication protocol.
This protocol is based on the continuous authentication scheme proposed by \citet{bours2012continuous}, which lacks privacy.
The authentication is performed in the encrypted domain, so the authentication server cannot learn anything about prestored templates, except the Boolean results and the key index $i$.
Our proposed protocol uses two types of cryptosystems as building-blocks:

\vspace{2mm}
\noindent \textbf{Homomorphic public key encryption algorithm.}
We assume a public key encryption algorithm (e.g., the Paillier cryptosystem) that supports following homomorphic property:
$E(m_1)\cdot E(m_2)= E(m_1+m_2)$.
For multiple identical ciphertexts, this property can be expressed as $E(m)^{k}= E(k \cdot m)$. For notation we an encrypted value is denote with $E$ such as $E(x)$ is the encryption of $x$ and the $(C^*)$ presents the encryption of distance indicator C.

\vspace{2mm}
\noindent \textbf{Privacy preserving comparison sub-protocol (PPCP).}
We use PPCP to compare the distance and the threshold in a privacy-preserving way, which takes one encrypted input $E(x)$ and an unencrypted input $y$, and determines whether $x>y$ without disclosing the values of $x$. 
These feature can be met by the private comparison protocol of \citet{damgaard:2007,damgaard:2008,damgaard:2009}. The other privacy-preserving protocol $PPCP^*$ takes two encrypted inputs $E(x)$, $E(y)$ and performs greater than comparison, this can be achieved by Veugen protocol \cite{veugen2012improving,veugen:2018}.

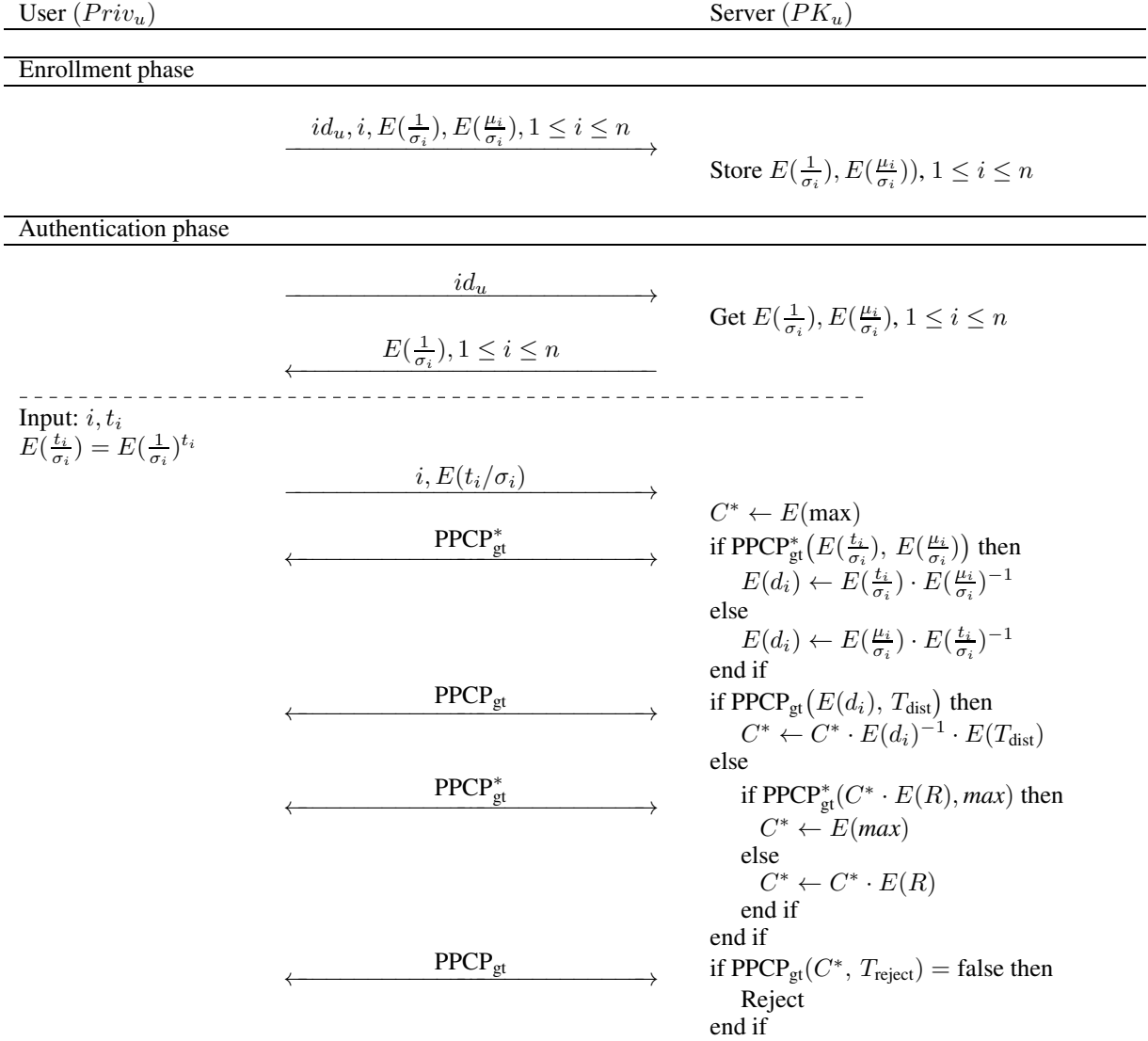
\begin{figure*}[t]
\begin{center}
\begin{tabular} {p{0.19\textwidth}   p{0.35\textwidth}  p{0.36\textwidth}}
User $(Priv_u)$ & &  Server $(PK_u)$   \\ \hline 
\\
\hline

Enrollment phase \\
\hline
  \\
									 		& $ \def\arraystretch{-1.0}
												 \begin{array}{c}
                                          id_{u}, i, E(\frac{1}{\sigma_i}), E(\frac{\mu_i}{\sigma_i}), 1 \leq i \leq n \\
                                          \parbox{5.3cm}{\rightarrowfill}
												\end{array} $  \\
											& & Store $E(\frac{1}{\sigma_i}), E(\frac{\mu_i}{\sigma_i}))$, $1 \leq i \leq n $ \\
\\
\hline
Authentication phase \\
\hline
\\
																		& $ \def\arraystretch{-1.0}
																														\begin{array}{c}
																				 id_{u} \\ \parbox{5.3cm}{\rightarrowfill}
																			\end{array} $
\\
											& &  Get $E(\frac{1}{\sigma_i}), E(\frac{\mu_i}{\sigma_i})$, $1 \leq i \leq n $  \\
											& $ \def\arraystretch{-1.0}
																				 \begin{array}{c}
																				 E(\frac{1}{\sigma_i}), 1 \leq i \leq n  \\ \parbox{5.3cm}{\leftarrowfill}
																			\end{array} $  \\
\begin{tikzpicture}
\draw [dashed] (0,0) -- (12,0);
\end{tikzpicture}
\\
Input: $i, t_i$	\\
$ E(\frac{t_i}{\sigma_i}) = E(\frac{1}{\sigma_i})^{t_i}   $   \\

		& $ \def\arraystretch{-1.0}
															\begin{array}{c} i,E(t_i/\sigma_i)  \\
															\parbox{5.3cm}{\rightarrowfill}
															\end{array} $
	\\																			
																					& &  $C^* \leftarrow E(\text{max})$ \\
																					 &  $ \def\arraystretch{-1.0}
                        	\begin{array}{c} \text{PPCP}^*_{\text{gt}}  \\
                          \parbox{5.3cm}{\leftarrowfill\hspace{-1mm}\rightarrowfill}
                        	\end{array} $      & if \text{PPCP}$^*_{\text{gt} } \big( E(\frac{t_i}{\sigma_i}), \, E(\frac{\mu_i}{\sigma_i}) \big)$ then \\
																					& & $\quad$ $E(d_i) \gets E(\frac{t_i}{\sigma_i})\cdot E(\frac{\mu_i}{\sigma_i})^{-1}$ \\
																					& & else \\
																					& & $\quad$ $E(d_i) \gets  E(\frac{\mu_i}{\sigma_i})\cdot E(\frac{t_i}{\sigma_i})^{-1}$  \\
                     & & end if \\
   &  $ \def\arraystretch{-1.0}
                        	\begin{array}{c} \text{PPCP}_{\text{gt}}  \\
                        \parbox{5.3cm}{\leftarrowfill\hspace{-1mm}\rightarrowfill}
                        	\end{array} $     & if PPCP${}_{\text{gt} } \big(E(d_i), \, T_{\text{dist}} \big)$ then \\
															  & & $\quad$ $C^* \gets C^* \cdot E(d_i)^{-1}\cdot E(T_{\text{dist}} )  $\\
																& & else \\
	 &  $ \def\arraystretch{-1.0}
                        	\begin{array}{c} \text{PPCP}^*_{\text{gt}}  \\
                          \parbox{5.3cm}{\leftarrowfill\hspace{-1mm}\rightarrowfill}
                        	\end{array} $      & $\quad$ if $\text{PPCP}^*_{\text{gt}}(C^* \cdot E(R), \text{\emph{max}})$ then \\
													&& $\quad\quad C^*\gets E(\text{\emph{max}}) $\\
														&&  $\quad$  else \\
														&& 		$\quad\quad C^*\gets C^* \cdot E(R) $ \\
															& & $\quad$ end if \\						
																& &  end if \\
   &  $ \def\arraystretch{-1.0}
                        	\begin{array}{c} \text{PPCP}_{\text{gt}}  \\
                        \parbox{5.3cm}{\leftarrowfill\hspace{-1mm}\rightarrowfill}
                        	\end{array} $        &   if $\text{PPCP}_{\text{gt}}(C^*, \, T_{\text{reject}}) = \text{false}$ then \\
																& & $\quad$ Reject \\
																& &  end if \\
\hline
\end{tabular}
\caption{Proposed privacy-preserving protocol for keystroke dynamics-based authentication}
\label{fig:proposed}
\end{center}
\end{figure*}

The proposed privacy-preserving continuous authentication protocol is presented in Figure~\ref{fig:proposed}.
It consists of the following three phases: Setup phase, enrollment phase, and authentication phase.
The detailed description of each phase is stated in the following:

\vspace{2mm}
\noindent \textbf{Enrollment phase.}
During the enrollment phase, the user registers himself to the server.
For this the user creates a key pair, and sends his public keys to the server.
The biometric features are collected at the user side.
We consider the following features are extracted from a keystroke: down-time ($t_i^{down}$), up-time ($t_i^{up}$) for every key $i$.
The time duration ($t_i$), the mean ($\mu_i$), and the standard deviation ($\sigma_i$) are computed in the similar manners as stated in Section~\ref{sec:ca}.
During the enrollment phase, the user encrypts the reference template $E(\frac{\mu_i}{\sigma_i}), E(\frac{1}{\sigma_i}), 1 \leq i \leq n$, for each key and sends encrypted template along with the user identity $id_u$ and the key index $i$ to the server.
The server stores $id_u$ and $E(\frac{\mu_i}{\sigma_i}), E(\frac{1}{\sigma_i}), 1 \leq i \leq n$, according to the index~$i$ of each key. Note that the user device does not store template locally.


\vspace{2mm}
\noindent \textbf{Authentication phase.}
During the authentication phase, the user initializes the protocol by sending the authentication request with his identity $id_u$ to the server.
The server searches for user identity $id_u$ and extracts them template that matches $id_u$.
Next, the server sends $E(\frac{1}{\sigma_i})$,  $1 \leq i \leq n$, to the user.

The remaining part is conducted each time that the user presses a key, which has index $i$.
The user computes the time duration $t_i=t_i^{up}-t_i^{down}$ of the pressed key, which is input to the homomorphic computation $E(\frac{t_i}{\sigma_i})= E(\frac{1}{\sigma_i})^{t_i}$ that is sent back to the server.

The server receives $E(\frac{t_i}{\sigma_i})$ and the server already holds reference template $E(\frac{\mu_i}{\sigma_i})$.
The server homomorphically computes one of the encrypted Scaled Manhattan Distance $E(d_i)$ between encrypted input and encrypted reference templates. 

In accordance with Eq.~\ref{eq:smd}, the Scaled Manhattan Distance needs to assign an absolute result\footnote{This is not included in the original paper}. 
The absolute result cannot be achieved directly under encryption. To get absolute result, we invoke a privacy-preserving greater than comparison protocol $(\text{PPCP}^*_{\text{gt}})$. This protocol determines the greater value between encrypted input and encrypted reference templates, then the $E(d_i)$ is computed in either of following way   
\begin{equation} \label{eq:encsimilarity1}
E(d_i)\gets E(\frac{t_i}{\sigma_i})\cdot E(\frac{\mu_i}{\sigma_i})^{-1}
\end{equation}
or 
\begin{equation} \label{eq:encsimilarity2}
E(d_i) \gets E(\frac{\mu_i}{\sigma_i}) \cdot  E(\frac{t_i}{\sigma_i})^{-1} 
\end{equation} 
This process always assigns positive value to $E(d_i)$ which is either achieved by Eq.~(\ref{eq:encsimilarity1}) or Eq.~(\ref{eq:encsimilarity2}).
 
As templates are encrypted with user public key, the server cannot find any information about the templates.
The encrypted distance $E(d_i)$ is compared to a predefined threshold $T_{\text{dist}}$ in privacy-preserving manners. 
The privacy-preserving comparison is explained in the following.

\subsection{Privacy-preserving comparison} \label{sec:comparison}

The presented protocol invokes a privacy-preserving comparison sub-protocol for the following tasks:
1)~To determine the greater value between $ E(\frac{t_i}{\sigma_i})$ and $E(\frac{\mu_i}{\sigma_i})$;
2)~compare the encrypted distance $E(d_i)$ and threshold $T_{\text{dist}}$ to decide whether to compute the privacy-preserving reward or the penalty functions;
3)~to compare the encrypted aggregated distance indicator $C^*$ with a preassigned maximum value w.r.t. the reward function; and
4)~to compare $C^*$ with the reject threshold value $T_{\text{reject}}$.


The reason for utilizing the privacy-preserving protocol is to hide the exact resultant values from the server and also from the malicious users. 

The server holds the encrypted SMD $E(d_i)$ and unencrypted threshold $T_{\text{dist}}$.
The server and the user invoke $\text{PPCP}_{\text{gt} } \big(E(d_i), \, T_{\text{dist}} \big)$ to check whether the distance is greater than the threshold.
The comparison protocol returns a Boolean result.
If true, the encrypted distance indicator $C^*$ is computed by the following penalty function:
\begin{equation}\label{eq:penalty}
\begin{split}
C^* &\gets C^* \cdot E(d_i)^{-1}\cdot E(T_{\text{dist}}) \\
   &= C^* \cdot E(-d_i)\cdot E(T_{\text{dist}}) 
    =  E(C-d_i+T_{\text{dist}})
\end{split}
\end{equation}
which decreases the plaintext value $C$ of $C^*$.

If false, the reward function is computed, which increments the plaintext value $C$ of $C^*$ as long as it is below the maximum value according to the Bours reward function:
\begin{equation}\label{eq:reward}
C \gets min(C+R, \text{max})
\end{equation}
In our protocol, this is realized by an if-block, where a privacy-preserving comparison is invoked for the third time:
\begin{equation*} 
\text{PPCP}^*_{\text{gt}}(C^* \cdot E(R), \text{\emph{max}})
\end{equation*}
where the $E(R)$ is the encrypted constant reward value. 
If the comparison is true, the $E(\emph{max})$ is assigned to $C^*$.
Otherwise, $C^* \gets C^* \cdot E(R) = E(C+R)$.

Lastly, a fourth privacy-preserving comparison $\text{PPCP}_{gt}(C^*, \, T_{\text{reject}})$ comparing the reject threshold with $(C^*)$.
If $C^*$ is below the reject threshold $(T_{\text{reject}})$, then the authentication fails and the user is rejected.


%% file: 7_analysis.tex
\section{\uppercase{analysis}} \label{sec:analysis}
In this section, we provide the correctness analysis, security analysis, and analysis of computation and communication complexity.

\subsection{Correctness analysis}
The correctness of our proposed protocol relies on additive homomorphic encryption properties.
The continuous authentication phase is entirely performed on encrypted templates. This section considers three kinds of computations performed in the encrypted domain: the correctness of the encrypted Scaled Manhattan Distance $E(d_i)$, the encrypted penalty function, and the encrypted reward function.

During the enrollment phase the user sends encrypted template $E(\frac{1}{\sigma_i}),\, E(\frac{\mu_i}{\sigma_i}), 1 \leq i \leq n$, to the server, and during the authentication phase the server receives encrypted input $E(\frac{t_i}{\sigma_i})= E(\frac{1}{\sigma_i})^{t_i}$.
The correctness proof of the encrypted Scaled Manhattan Distance $E(d_i)$ in Eqs.~(\ref{eq:encsimilarity1}, \ref{eq:encsimilarity2}) can be verified by the following equation:
\begin{equation*}
\begin{split}
 E(d_i)=&E(\frac{t_i}{\sigma_i})\cdot E(\frac{\mu_i}{\sigma_i})^{-1} = E(\frac{t_i-\mu_i}{\sigma_i})
\end{split}
\end{equation*}
when $E(\frac{\mu_i}{\sigma_i})$ gets greater than $E(\frac{t_i}{\sigma_i})$, then $E(d_i)$ can be achieved as  
\begin{equation*}
\begin{split}
 E(d_i)=& E(\frac{\mu_i}{\sigma_i})\cdot E(\frac{t_i}{\sigma_i})^{-1} = E(\frac{\mu_i-t_i}{\sigma_i})
\end{split}
\end{equation*}
Where the Scaled Manhattan Distance, stated in Eq.~\ref{eq:smd}.

The aggregated distance indicator $C^*$ is computed in the form of penalty and reward functions in Eqs.~(\ref{eq:penalty}, \ref{eq:reward}).
The correctness proof of the encrypted penalty function is
\begin{equation*} 
\begin{split}
C^*= C^* \cdot E(d_i)^{-1}\cdot E(T_{\text{dist}}) = E(C-d_i+T_{dist})\\
\end{split}
\end{equation*}
and the correctness proof of the encrypted reward function is
\begin{equation*} 
\begin{split}
C^*=&\text{min}(C^* \cdot E(R), E(\text{\emph{max}}))=E(C + R), E(\text{\emph{max}})
\end{split}
\end{equation*}
where $C^*$ is the encryption of $C$, $E(R)$ is the encryption of reward value $R$, and $E(\text{\emph{max}})$ is the encryption of maximum value.

\subsection{Security analysis}
Our generic protocol relies on the security properties of additive homomorphic cryptosystems, e.g., \citet{Paillier:99}.
As noted, the privacy requirement is that the stored reference templates and input templates are protected so that the server cannot learn anything about them.
Our protocol achieves the privacy in the following ways:
1) The reference template is stored in the encrypted form,
2) Scaled Manhattan Distance is computed on the encrypted input and encrypted reference templates),
3) the comparison between the threshold and the result is made in a privacy-preserving way,
4) the aggregated indicator $C$ is computed in the encrypted form and,
5) the final comparison is also made in a privacy-preserving way using PPCP.
The server or malicious insider on the server cannot see any additional information about biometric templates. Hence, this protocol is a fully privacy-preserving protocol.

\begin{table*}[t]
\caption{Complexity comparison}
\label{tb6:comparison}
\centering
\begin{tabular}{l r r r  }
\toprule
Protocol
 & Rounds 
 & Transmitted encryptions  
 & Sub-protocols invocations
\\ \midrule
\cite{govindarajan:2013}   &  4 &  $N+1$   & $4 N$\\
\cite{balagani:2018} 	    &  5 &  $2 N+1$ & $5 N$\\
 \cite{wei:2020}           &  3 &  $3 N$	  & 0  \\
Our protocol                &  5 &	$ N+4$	&  4\\
\bottomrule
\end{tabular}	
\end{table*}

\subsection{Performance analysis}
The performance of a protocol can be determined by analyzing the computation and communication complexities. In this context, we analyzed the number of rounds to complete the authentication decision, the number of transmitted encryptions, and the number of invocations of sub-protocols for privacy-preserving comparison. We compared our protocol with other protocols that have been proposed only for behavioral biometrics limited to touch-dynamics or keystroke dynamics. Context-aware authentication modes, such as authentication based on GPS data, web-histories, IP addresses; e.g., \citet{domingo:2015,shahandashti:2015} are not considered for comparison.

Our protocol performs continuous authentication in five rounds. Each round contains only one encryption, except for the first round which is performed only once.
Our protocol invokes a sub-protocol for privacy-preserving comparison four times, where the second time PPCP$_{\text{gt}}$ is only invoked when $d_i<T_{\text{dist}}$.
Our protocol is based on only one sampled input that compares only two integers (resultant value and a threshold).

The \citeauthor{govindarajan:2013} protocol transmits $N+1$ encryptions during the authentication phase. The first round transmits $N$ encrypted elements and the second round transmits only one encrypted element. The authentication decision is completed by four times invoking the privacy-preserving comparison protocol. Each time the sub-protocol compares the series of $N$ encrypted elements of a feature vector.
These sub-protocols are based on the \citet{erkin:2009} protocol, and the \citeauthor{erkin:2009}\space utilizes the \citet{damgaard:2007,damgaard:2009} comparison protocol.
As \citeauthor{govindarajan:2013}\space invoked a sub-protocol four times for $N$ samples, where one comparison is completed in three rounds. Their protocol takes total $12 \times N$ rounds to complete an authentication decision.

The \citet{balagani:2018} protocol completes authentication in five rounds, and they transmitted $2 N+1$ encryptions. Moreover, they five times invoked sub-protocols to complete one decision.
They also utilized the \citeauthor{erkin:2009}\space protocol for privacy-preserving comparison.
Their protocol completes authentication decision in $15 \times N$ total rounds.

The Wei et al. \citet{wei:2020} protocol utilizes Paillier cryptosystem and completes an authentication decision in three rounds.
They transmit $3 N$ encrypted vectors, where each vector contained $N$ encrypted elements in a vector.

In comparison to \citet{govindarajan:2013,balagani:2018,wei:2020}, our protocol is efficient in terms of computation cost, other protocols compute and transmit $N$ the encrypted elements in each round, whereas our protocol transmits only one encrypted element. The comparison of $N$ encrypted elements in each round makes them very inefficient even for periodic authentication. Therefore, our protocol is very efficient for continuous authentication.

The comparison is presented in Table~\ref{tb6:comparison}, where the number of rounds and transmitted encryptions are counted without including the sub-protocols.

\subsection{Implementation}

To determine the performance of our proposed protocols, we implemented the authentication protocol on Intel(R) Core(TM) i5-7440 HQ CPU @ 2.80GHz, 32 GB RAM in Python 3.10 (Jupyter Notebook, Anaconda3).
We use two libraries for our implementation; the homomorphic encryption library \citep{PHE:lib} and a library for the secure comparison protocol, developed by \citet{TNO:lib}.
To evaluate the computation cost, we measured execution time to complete homomorphic operations, and to complete privacy-preserving comparisons. 

The execution time of continuous authentication is presented in Table~\ref{tb7:executiontime}. We utilize Veugen protocol \cite{veugen2012improving,veugen:2018} as sub-protocol for privacy-preserving comparison that is a bit-wise comparison protocol. Note that Veugen protocol is the improvement of DGK comparison protocol (\citet{damgaard:2007,damgaard:2009}).
The key-size is represented by $k$ and $l$ represents the bit-length of input numbers of Veugen protocol \cite{veugen2012improving,veugen:2018}.   
The performance is determined by setting security parameter $k$ into different sizes $(k=512$ to $1536)$, and used different bit-length of input numbers $(l=5$ - $l=10)$-bits. The execution time presented in Table~in Table~\ref{tb7:executiontime} is calculated in milliseconds (ms)\footnote{This implementation invokes $4T_{PPCP}$, whereas the original paper three times invokes the PPCP, and due to page restriction Table~\ref{tb7:executiontime} is not presented in the original paper}.

\begin{table*}
\caption{Execution time of proposed protocol}
\label{tb7:executiontime}
\centering
\begin{tabular}{l r r r  }
\toprule
$k$
 &$l=4$ 
 &$l=7$ 
 &$l=10$
\\ \midrule
 512 &$\approx80$ms&$\approx125$ms&$\approx195$ms\\
 768 &$\approx255$ms&$\approx$390ms&$\approx500$ms\\
1024 &$\approx540$ms&$\approx750$ms&$\approx1006$ms\\
1536 &$\approx1850$ms&$\approx2503$ms&$\approx3201$ms\\
\bottomrule
\end{tabular}	
\end{table*}



%% file: 8_conclusion.tex
\section{\uppercase{conclusions and future work}} \label{sec:conclusions}
Continuous authentication strengthens the security by monitoring the user behavioral features but causes privacy concerns when behavioral features are transmitted to the authentication server.
In this paper, we have presented a new generic privacy-preserving keystroke dynamics-based continuous authentication protocol. We have proposed a simple and efficient privacy-preserving protocol utilizing homomorphic encryption properties as building blocks. Our protocol does not reveal any information about the biometric templates or the resultant outputs.
This protocol provides privacy against the honest-but curious server.
To the best of our knowledge, our protocol is the first one to offer privacy-preserving continuous authentication. Moreover, our protocol provides efficient performance compared to the literature.

Multimodal behavioral biometric-based continuous authentication may offer more security than a unimodal authentication mechanism.
We will consider multimodal continuous authentication using behavioral biometrics in the future. 
This will be achieved by combining keystroke dynamics with touch dynamics.
Continuous authentication requires efficient performance in terms of communication and computation costs, and our future research will focus on lightweight encryption techniques to reduce communication overhead.

%% file: 0_PP_keystroke.bbl
\begin{thebibliography}{}

\bibitem[Baig. and Eskeland., 2022]{secrypt22}
Baig., A. and Eskeland., S. (2022).
\newblock A generic privacy-preserving protocol for keystroke dynamics-based
  continuous authentication.
\newblock In {\em Proceedings of the 19th International Conference on Security
  and Cryptography - SECRYPT,}, pages 491--498. INSTICC, SciTePress.

\bibitem[Baig and Eskeland, 2021]{baig2021security}
Baig, A.~F. and Eskeland, S. (2021).
\newblock Security, privacy, and usability in continuous authentication: A
  survey.
\newblock {\em Sensors}, 21(17):5967.

\bibitem[Balagani et~al., 2018]{balagani:2018}
Balagani, K.~S., Gasti, P., Elliott, A., Richardson, A., and O’Neal, M.
  (2018).
\newblock The impact of application context on privacy and performance of
  keystroke authentication systems.
\newblock {\em Journal of Computer Security}, 26(4):543--556.

\bibitem[Boldyreva et~al., 2009]{boldyreva:2009}
Boldyreva, A., Chenette, N., Lee, Y., and O’neill, A. (2009).
\newblock Order-preserving symmetric encryption.
\newblock In {\em Annual International Conference on the Theory and
  Applications of Cryptographic Techniques}, pages 224--241. Springer.

\bibitem[Bours, 2012]{bours2012continuous}
Bours, P. (2012).
\newblock Continuous keystroke dynamics: A different perspective towards
  biometric evaluation.
\newblock {\em Information Security Technical Report}, 17(1-2):36--43.

\bibitem[Brizan et~al., 2015]{brizan2015utilizing}
Brizan, D.~G., Goodkind, A., Koch, P., Balagani, K., Phoha, V.~V., and
  Rosenberg, A. (2015).
\newblock Utilizing linguistically enhanced keystroke dynamics to predict
  typist cognition and demographics.
\newblock {\em International Journal of Human-Computer Studies}, 82:57--68.

\bibitem[Damg{\aa}rd et~al., 2007]{damgaard:2007}
Damg{\aa}rd, I., Geisler, M., and Kr{\o}igaard, M. (2007).
\newblock Efficient and secure comparison for on-line auctions.
\newblock In {\em Australasian conference on information security and privacy},
  pages 416--430. Springer.

\bibitem[Damgard et~al., 2008]{damgaard:2008}
Damgard, I., Geisler, M., and Kroigard, M. (2008).
\newblock Homomorphic encryption and secure comparison.
\newblock {\em International Journal of Applied Cryptography}, 1(1):22--31.

\bibitem[Damgard et~al., 2009]{damgaard:2009}
Damgard, I., Geisler, M., and Kroigard, M. (2009).
\newblock A correction to'efficient and secure comparison for on-line
  auctions'.
\newblock {\em International Journal of Applied Cryptography}, 1(4):323--324.

\bibitem[Dhakal et~al., 2018]{dhakal2018observations}
Dhakal, V., Feit, A.~M., Kristensson, P.~O., and Oulasvirta, A. (2018).
\newblock Observations on typing from 136 million keystrokes.
\newblock In {\em Proceedings of the 2018 CHI Conference on Human Factors in
  Computing Systems}, pages 1--12.

\bibitem[Domingo-Ferrer et~al., 2015]{domingo:2015}
Domingo-Ferrer, J., Wu, Q., and Blanco-Justicia, A. (2015).
\newblock Flexible and robust privacy-preserving implicit authentication.
\newblock In {\em IFIP International Information Security and Privacy
  Conference}, pages 18--34. Springer.

\bibitem[Erkin et~al., 2009]{erkin:2009}
Erkin, Z., Franz, M., Guajardo, J., Katzenbeisser, S., Lagendijk, I., and Toft,
  T. (2009).
\newblock Privacy-preserving face recognition.
\newblock In {\em International symposium on privacy enhancing technologies
  symposium}, pages 235--253. Springer.

\bibitem[Freedman et~al., 2004]{freedman:2004}
Freedman, M.~J., Nissim, K., and Pinkas, B. (2004).
\newblock Efficient private matching and set intersection.
\newblock In {\em International conference on the theory and applications of
  cryptographic techniques}, pages 1--19. Springer.

\bibitem[Govindarajan et~al., 2013]{govindarajan:2013}
Govindarajan, S., Gasti, P., and Balagani, K.~S. (2013).
\newblock Secure privacy-preserving protocols for outsourcing continuous
  authentication of smartphone users with touch data.
\newblock In {\em 2013 IEEE Sixth International Conference on Biometrics:
  Theory, Applications and Systems (BTAS)}, pages 1--8. IEEE.

\bibitem[Paillier, 1999]{Paillier:99}
Paillier, P. (1999).
\newblock Public-key cryptosystems based on composite degree residuosity
  classes.
\newblock In {\em International conference on the theory and applications of
  cryptographic techniques}, pages 223--238. Springer.

\bibitem[{Python-paillier.readthedocs.io}, 2016]{PHE:lib}
{Python-paillier.readthedocs.io} (2016).
\newblock {Python library for Partially Homomorphic Encryption}.
\newblock {\url{https://python-paillier.readthedocs.io/en/develop/index.html}}.
\newblock {[Accessed 11.05.2022]}.

\bibitem[Safa et~al., 2014]{safa:2014}
Safa, N.~A., Safavi-Naini, R., and Shahandashti, S.~F. (2014).
\newblock Privacy-preserving implicit authentication.
\newblock In {\em IFIP International Information Security Conference}, pages
  471--484. Springer.

\bibitem[Shahandashti et~al., 2015]{shahandashti:2015}
Shahandashti, S.~F., Safavi-Naini, R., and Safa, N.~A. (2015).
\newblock Reconciling user privacy and implicit authentication for mobile
  devices.
\newblock {\em Computers \& Security}, 53:215--233.

\bibitem[{TNO MPC Lab}, 2021]{TNO:lib}
{TNO MPC Lab} (2021).
\newblock https:.
\newblock \url{//pypi.org/project/tno.mpc.protocols.secure-comparison/}.
\newblock {[Accessed 11.05.2022]}.

\bibitem[Veugen, 2012]{veugen2012improving}
Veugen, T. (2012).
\newblock Improving the dgk comparison protocol.
\newblock In {\em 2012 IEEE International Workshop on Information Forensics and
  Security (WIFS)}, pages 49--54. IEEE.

\bibitem[Veugen, 2018]{veugen:2018}
Veugen, T. (2018).
\newblock Correction to" improving the dgk comparison protocol".
\newblock {\em Cryptology ePrint Archive}.

\bibitem[Wei et~al., 2020]{wei:2020}
Wei, F., Vijayakumar, P., Kumar, N., Zhang, R., and Cheng, Q. (2020).
\newblock Privacy-preserving implicit authentication protocol using cosine
  similarity for internet of things.
\newblock {\em IEEE Internet of Things Journal}, 8(7):5599--5606.

\bibitem[Xiaofeng et~al., 2019]{xiaofeng2019continuous}
Xiaofeng, L., Shengfei, Z., and Shengwei, Y. (2019).
\newblock Continuous authentication by free-text keystroke based on cnn plus
  rnn.
\newblock {\em Procedia computer science}, 147:314--318.

\end{thebibliography}
